\documentclass{ws-ijmpa}

\begin{document}

\markboth{Wei Chao, Shu Luo, Zhi-zhong Xing and Shun Zhou}
{TeV-scale Type-II Seesaw Models and Possible Collider Signatures}

%
%

\title{TeV-scale Type-II Seesaw Models and Possible Collider Signatures}

\author{ Wei Chao$^*$, Shu Luo, Zhi-zhong Xing and Shun Zhou}

\address{Institute of High Energy Physics, Chinese Academy of
Sciences, Beijing 100049, China\\$^*$E-mail: chaowei@ihep.ac.cn}

\maketitle

\begin{history}
\received{\today}
\end{history}

\begin{abstract}
A natural extension of the standard model to accommodate massive
neutrinos is to introduce one Higgs triplet and three right-handed
Majorana neutrinos, leading to a $6\times6 $ neutrino mass matrix.
We show that three light Majorana neutrinos (i.e., the mass
eigenstates of $\nu^{}_e$, $\nu^{}_\mu$ and $\nu^{}_\tau$) are
exactly massless, if and only if $M^{}_{\rm L} = M^{}_{\rm D}
M^{-1}_{\rm R} M^T_{\rm D}$ exactly holds in this seesaw model. We
propose three simple Type-II seesaw scenarios with broken $A^{}_4
\times U(1)^{}_{\rm X}$ flavor symmetry to interpret the observed
neutrino mass spectrum and neutrino mixing pattern. Such a TeV-scale
neutrino model can be tested in two complementary ways: (1)
searching for possible collider signatures of lepton number
violation induced by the right-handed Majorana neutrinos and
doubly-charged Higgs particles; and (2) searching for possible
consequences of unitarity violation of the $3\times 3$ neutrino
mixing matrix in the future long-baseline neutrino oscillation
experiments.

\end{abstract}

\ccode{PACS numbers: 11.30.Fs, 14.60.Pq, 14.60.St}

\vspace{0.2cm}

The discovery of neutrino oscillations has confirmed the theoretical
expectation that neutrinos are massive and lepton flavors are mixed,
providing the first evidence for physics beyond the standard model
(SM) in particle physics. At present, the seesaw mechanism is the
most natural idea to understand the smallness of neutrino masses.
However, how to test it has been an open question.

We shall talk about the Type-II seesaw mechanism,\cite{typeII} which
extends the SM with three righ-handed Majorana neutrinos and one
Higgs triplet. The effective mass matrix for three light neutrinos
is given by $M^{}_\nu \approx M^{}_{\rm L} - M^{}_{\rm D}
M^{-1}_{\rm R} M^T_{\rm D}$. To test the Type-II seesaw model, one
possible way is to lower the mass scales of both the heavy neutrinos
and the Higgs triplet to the TeV level and allow the Yukawa coupling
matrices $Y^{}_{\nu}$ and $Y_\Delta^{}$ to be of ${\cal O}(10^{-3})$
up to ${\cal O}(1)$. Then we may search for the
lepton-number-violating signals induced by those heavy particles at
the Large Hadron Collider (LHC). In order to generate sufficiently
small neutrino masses in this kind of TeV-scale seesaw scenarios,
the key point is to adjust the textures of $M_{\rm L}^{}$, $M_{\rm
D}^{}$ and $M_{\rm R}^{}$ to guarantee $M_{\rm L}^{} = M_{\rm
D}^{}M_{\rm R}^{-1} M_{\rm D}^T$ in the leading-order approximation.
Then tiny but non-vanishing neutrino masses can be ascribed to
slight perturbations or radiative corrections to $M_{\rm L}^{}$ and
$M_{\rm D}^{}M_{\rm R}^{-1}M_{\rm D}^T$. We shall prove a no-go
theorem: the masses of light Majorana neutrinos are exactly
vanishing at the tree level if and only if the global cancellation
$M_{\rm L}^{}-M_{\rm D}^{}M_{\rm R}^{-1}M_{\rm D}^T=0$ exactly holds
in generic Type-II seesaw scenarios. Therefore, a feasible way to
obtain both tiny neutrino masses and appreciable collider signatures
is to allow for an incomplete cancellation between $M^{}_{\rm L}$
and $M^{}_{\rm D} M^{-1}_{\rm R} M^T_{\rm D}$ terms. We shall
propose three simple Type-II seesaw scenarios at the TeV scale by
taking into account the $A^{}_4 \times U(1)^{}_{\rm X}$ flavor
symmetry and its breaking mechanism, from which the observed
neutrino mass spectrum and neutrino mixing pattern can be achieved.
We shall also discuss two interesting consequences of this model:
(1) possible unitarity violation of the $3\times 3$ neutrino mixing
matrix, which can be searched for in the future long-baseline
neutrino oscillation experiments; and (2) possible signatures of
lepton number violation induced by the right-handed Majorana
neutrinos and doubly-charged Higgs particles, which can be searched
for at colliders.

We regularize our notations and conventions by reviewing some basics
of the Type-II seesaw mechanism. After spontaneous symmetry
breaking, the lepton mass terms turn out to be
\begin{eqnarray}
-{\cal L}_{\rm mass} = \overline{E^{}_{\rm L}} M^{}_l E^{}_{\rm R} +
\frac{1}{2} \overline{\left( \nu^{}_{\rm L} ~N^c_{\rm R}\right)}
\left(
\begin{matrix}
 M^{}_{\rm L} & M^{}_{\rm D} \cr M^T_{\rm D} &
M^{}_{\rm R}
\end{matrix}
\right) \left(
\begin{matrix}
\nu^c_{\rm L} \cr N^{}_{\rm R}
\end{matrix}
\right) + {\rm h.c.} \; .
\end{eqnarray}
The overall $6\times 6$ neutrino mass matrix in ${\cal L}^{}_{\rm
mass}$, denoted as ${\cal M}$, can be diagonalized by the unitary
transformation ${\cal U}^\dagger {\cal M} {\cal U}^* = \widehat{\cal
M}$; or explicitly,
\begin{eqnarray}
\left(
\begin{matrix}
V & R \cr S & U
\end{matrix}
\right)^\dagger \left(
\begin{matrix}
 M^{}_{\rm L} & M^{}_{\rm D}
\cr M^T_{\rm D} & M^{}_{\rm R}
\end{matrix}
\right) \left(
\begin{matrix}
V & R \cr S & U
\end{matrix}
\right)^*  = \left( \begin{matrix} \widehat{M}^{}_\nu & {\bf 0} \cr
{\bf 0} & \widehat{M}^{}_{\rm N}\end{matrix}\right) \; ,
\end{eqnarray}
where $\widehat{M}^{}_\nu = {\rm Diag}\{m^{}_1, m^{}_2, m^{}_3\}$
and $\widehat{M}^{}_{\rm N} = {\rm Diag}\{M^{}_1, M^{}_2, M^{}_3\}$
with $m^{}_i$ and $M^{}_i$ (for $i=1, 2, 3$) being the light and
heavy Majorana neutrino masses, respectively.

In the basis where the flavor eigenstates of three charged leptons
are identified with their mass eigenstates, the standard
charged-current interactions between $\nu^{}_\alpha$ and $\alpha$
(for $\alpha = e, \mu, \tau$) turn out to be
\begin{eqnarray}
-{\cal L}^{}_{\rm cc} = \frac{g}{\sqrt{2}} \left[
\overline{\left(e~~ \mu~~ \tau\right)^{}_{\rm L}} V \gamma^\mu
\left( \begin{matrix}\nu^{}_1 \cr \nu^{}_2 \cr \nu^{}_3\end{matrix}
\right)^{}_{\rm L} W^-_{\mu} + \overline{\left(e~~ \mu~~
\tau\right)^{}_{\rm L}} R \gamma^\mu \left( \begin{matrix}N^{}_1 \cr
N^{}_2 \cr N^{}_3\end{matrix} \right)^{}_{\rm L} W^-_\mu \right] +
{\rm h.c.} \; . ~~~
\end{eqnarray}
In a TeV-scale Type-II seesaw model with the complete cancellation
between $M_{\rm L}^{}$ and $M_{\rm D}^{} M_{\rm R}^{-1} M_{\rm
D}^T$, it seems that tiny neutrino masses could be generated by the
sub-leading terms of $M_\nu^{}$. However, this idea does not work
because of the following no-go theorem:\cite{zhluo} {\it If and only
if the relationship $M^{}_{\rm L} = M^{}_{\rm D} M^{-1}_{\rm R}
M^T_{\rm D}$ is exactly satisfied in Type-II seesaw models, then
three light Majorana neutrinos must be exactly massless.}

Let us prove the theorem given above. The key point is to replace
$M^{}_{\rm L}$ with $M^{}_{\rm D} M^{-1}_{\rm R} M^T_{\rm D}$ in
${\cal M}$, and then the first row of $\cal M$ can simply be
obtained from the second row of $\cal M$ multiplied by $M^{}_{\rm D}
M^{-1}_{\rm R}$ on the left. It follows that the rank of ${\cal M}$
must be equal to three (i.e., the rank of $M^{}_{\rm R}$), implying
that three light Majorana neutrinos must be exactly massless under
the condition of $M^{}_{\rm L} = M^{}_{\rm D} M^{-1}_{\rm R}
M^T_{\rm D}$.

To simultaneously achieve tiny neutrino masses and large neutrino
mixing angles, we impose the $A^{}_4 \times U(1)^{}_{\rm X}$ flavor
symmetry\cite{A4} on the Type-II seesaw Lagrangian. In this case,
the assignments of relevant lepton and scalar fields with respect to
the symmetry group $SU(2)^{}_{\rm L} \times U(1)^{}_{\rm Y} \otimes
A^{}_4 \times U(1)^{}_{\rm X}$ are:  $l^{}_{\rm L} \sim (2, -1)
\otimes (\underline{3}, 1),\; E^{}_{\rm R} \sim (1, -2) \otimes
(\underline{1}, 1), \; E^{'}_{\rm R} \sim (1, -2) \otimes
(\underline{1}^\prime, 1) ,\; E^{''}_{\rm R} \sim (1, -2) \otimes
(\underline{1}^{\prime \prime}, 1)  ,\; \phi \sim (2, -1) \otimes
(\underline{1}, 1),\; \Phi \sim (2, -1) \otimes (\underline{3},
0),\; \chi \sim (1, 0) \otimes (\underline{3}, 1) , \; \Delta \sim
(3, -2) \otimes (\underline{1}, 2),\; N^{}_{\rm R} \sim (1, 0)
\otimes (\underline{3}, 0)  $ and $\Sigma \sim (3, -2) \otimes
(\underline{3}, 0), $ where several triplet scalars have been
introduced. Given $SU(2)^{}_{\rm L} \times U(1)^{}_{\rm Y} \otimes
A^{}_4 \times U(1)^{}_{\rm X}$ invariance, the Lagrangian
responsible for lepton masses reads
\begin{eqnarray}
- {\cal L}^{}_{\rm Y} = \sum_{\alpha}y^{\alpha}_e
(\overline{l^{}_{\rm L}} \tilde{\Phi})^{}_{\underline{\alpha}}
E^{\alpha}_{\rm R}+ \frac{y^{}_\Delta}{2} \overline{l^{}_{\rm L}}
i\sigma^{}_2 \Delta l^c_{\rm L} + \frac{m^{}_{\rm R}}{2}
(\overline{N^c_{\rm R}} N^{}_{\rm R})^{}_{\underline{1}} + y^{}_\nu
(\overline{l^{}_{\rm L}}N^{}_{\rm R})^{}_{\underline{1}} \phi + {\rm
h.c.} \; . ~~
\end{eqnarray}
After spontaneous symmetry breaking, the overall neutrino mass
matrix $\cal M$ is determined by its three $3\times 3$ sub-matrices:
$M^{}_{\rm L} = m^{}_{\rm L} \cdot {\bf 1}$, $M^{}_{\rm D} =
m^{}_{\rm D}\cdot {\bf 1}$ and $M^{}_{\rm R} = m^{}_{\rm R}\cdot
{\bf 1}$. In the assumption of $\langle \Phi^{}_1 \rangle = \langle
\Phi^{}_2 \rangle = \langle \Phi^{}_3 \rangle$, the charged-lepton
mass matrix can be written as $M^{}_l = U^{}_l \widehat{M}^{}_l$,
where $\widehat{M}^{}_l = {\rm Diag} \{m^{}_e, m^{}_\mu, m^{}_\tau
\} = \sqrt{3} \langle \Phi^{}_i \rangle {\rm Diag}\{y^{}_e,
y^{\prime}_e, y^{\prime \prime}_e \}$ and $U_l^{}$ can be found in
Ref. 2. It is quite obvious that $m^{}_{\rm L} = m^2_{\rm
D}/m^{}_{\rm R}$ will lead to $M^{}_{\rm L} = M^{}_{\rm D}
M^{-1}_{\rm R} M^T_{\rm D}$. According to the no-go theorem, this
complete cancellation makes light neutrino masses exactly vanishing.
To obtain the realistic neutrino mass spectrum and lepton flavor
mixing pattern,\cite{tribi} we may introduce an incomplete
cancellation between $M^{}_{\rm L}$ and $M^{}_{\rm D} M^{-1}_{\rm R}
M^T_{\rm D}$ terms by breaking the flavor symmetry $U(1)^{}_{\rm X}$
explicitly to $Z^{}_2$. We may assign the proper $Z^{}_2$ parity to
produce slight perturbations to the neutrino mass terms. There are
three simple possibilities:\cite{zhluo} (1) perturbations to
$M^{}_{\rm L}$, i.e., $l^{}_{\rm L}$, $E^{}_{\rm R}$,
$E^{\prime}_{\rm R}$, $E^{\prime \prime}_{\rm R}$, $\chi$ and $\phi$
are odd under the $Z^{}_2$ transformation, while the other fields
are even under the same transformation; (2) perturbations to
$M^{}_{\rm D}$, i.e., $l^{}_{\rm L}$, $\chi$, $\Sigma$ and $\phi$
are odd under the $Z^{}_2$ transformation, while the other fields
are even under the same transformation; and (3) perturbations to
$M^{}_{\rm R}$, i.e., $l^{}_{\rm L}$, $E^{}_{\rm R}$,
$E^{\prime}_{\rm R}$, $E^{\prime \prime}_{\rm R}$, $\Sigma$ and
$\phi$ are odd under the $Z^{}_2$ transformation, while the other
fields are even under the same transformation. Each of them is
compatible with current neutrino oscillation data. A more general
approach should include the perturbations to $M_{\rm L}^{}$, $M_{\rm
D}^{}$ and $M_{\rm R}^{}$ together. Then the Type-II seesaw formula
can be re-expressed as
\begin{eqnarray}
M^{}_\nu \approx \delta M + \delta M^{}_{\rm L} + \tilde{M}^{}_{\rm
D} \tilde{M}^{-1}_{\rm R} \delta M^{}_{\rm R} \tilde{M}^{-1}_{\rm R}
\tilde{M}^T_{\rm D}- \tilde{M}^{}_{\rm D} \tilde{M}^{-1}_{\rm R}
(\delta M^{}_{\rm D})^T - \delta M^{}_{\rm D} \tilde{M}^{-1}_{\rm R}
\tilde{M}^T_{\rm D} \; , ~~
\end{eqnarray}
where $\delta M_{\rm L, D, R}^{}$ is the perturbation term of
$M_{\rm L, D, R}^{}$, and $\delta M$ denotes the residue of the
incomplete cancellation between $M_{\rm L}^{}$ and $M_{\rm
D}^{}M_{\rm R}^{-1}M_{\rm D}^T$ terms.

Now we proceed to discuss the unitarity violation and collider
signatures in the Type-II seesaw model. The non-unitarity of the
lepton flavor mixing matrix $V$ is actually a common feature of the
seesaw models, as one can easily see from $VV^\dagger = {\bf 1} -
RR^\dagger \neq {\bf 1}$. In practice, one may resort to a recursive
expansion\cite{Grimus} of $M_\nu^{}$ in powers of $M_{\rm
D}^{}M_{\rm R}^{-1}$ and then obtain $\xi \equiv RR^\dagger \approx
{U^\dagger_l}^2 M^{}_{\rm D} M^{-1}_{\rm R} (M^{-1}_{\rm R} M^T_{\rm
D})^* {U^{}_l}^2$, where $U^{}_l$ is the unitary matrix used to
diagonalize $M^{}_l$. Note that $\xi $ is in general complex and may
give rise to some additional CP-violating effects in neutrino
oscillations.\cite{CPV} In our specific scenarios discussed above,
however, we find $\xi \approx m_{\rm D}^2/m_{\rm R}^2\cdot{\bf 1}$.
Thus there is almost no extra CP violation induced by the unitarity
violation of V. Translating the numerical results of Refs. 6-8 into
the restriction on $\xi$ in our language, we obtain
\begin{eqnarray}
\left|\xi\right| = \left(\begin{matrix}|\xi^{}_{ee}| < 1.1 \cdot
10^{-2} & |\xi^{}_{e\mu}| < 7.0 \cdot 10^{-5} & |\xi^{}_{e\tau}| <
1.6 \cdot 10^{-2} \cr |\xi^{}_{\mu e}| < 7.0 \cdot 10^{-5} &
|\xi^{}_{\mu\mu}| < 1.0 \cdot 10^{-2} & |\xi^{}_{\mu \tau}| < 1.0
\cdot 10^{-2} \cr |\xi^{}_{\tau e}| < 1.6 \cdot 10^{-2} &
|\xi^{}_{\tau \mu}| < 1.0 \cdot 10^{-2} & |\xi^{}_{\tau\tau}| < 1.0
\cdot 10^{-2} \cr\end{matrix}\right) \;
\end{eqnarray}
at the $90\%$ confidence level.

A direct test of the seesaw mechanism requires the unambiguous
observation of heavy Majorana neutrinos. The clearest signature
induced by $N^{}_i$ should be the lepton-number-violating process
$pp \to W^\pm \to \mu^\pm N \to \mu^\pm \mu^\pm jj$ at the LHC.
\cite{sen} It can be resonantly enhanced due to the on-shell
production of heavy Majorana neutrinos. We feel that the discovery
of heavy majorana neutrinos with  $M^{}_i \sim {\cal O}(10^2)~{\rm
GeV}$ to ${\cal O}(1)~{\rm TeV}$ and $\xi^{}_{\mu \mu} \sim {\cal
O}(10^{-3})$ to ${\cal O}(10^{-2})$ is possible. For the doubly
charged scalars existing in the Type-II seesaw model, one may
concentrate on the pair production in the Drell-Yan process $q
\bar{q} \to \gamma^*/ Z^* \to \Delta^{\pm \pm} \Delta^{\mp
\mp}$\cite{pair} and the subsequent decays $\Delta^{\pm \pm} \to
W^{\pm} W^{\pm}$ or $\Delta^{\pm \pm} \to l^{\pm} l^{\pm}$. The
doubly charged scalars can be observed at the LHC with a branching
fraction $\sim 50\%$ up to the mass range of $800~{\rm GeV}$ to
$1~{\rm TeV}$. In our model, the choice of $y^{}_\Delta \sim {\cal
O}(1)$ and $\langle \Delta \rangle \sim 1~{\rm GeV}$ will extend the
above mass range for the doubly-charged scalars.

In short, the main concern of this talk is the experimental
testability of the seesaw mechanism in the era of the LHC and
precision neutrino experiments. We find that the naturalness of the
Type-II seesaw mechanism might be partly lost at the TeV scale, just
like that of the Type-I seesaw mechanism at this energy
region.\cite{Smirnov} More theoretical and phenomenological effort
is desirable in order to bridge the gap between light and heavy
Majorana neutrinos.


\begin{thebibliography}{0}
\bibitem{typeII} J. Schechter and J. W. F. Valle, \emph{Phys. Rev. D}
{\bf 22}, 2227 (1980); T. P. Cheng and L. F. Li, \emph{Phys. Rev. D}
{\bf 22}, 2860 (1980); M. Magg and C. Wetterich, \emph{Phys. Lett. B
}{\bf 94}, 61 (1980).

\bibitem{zhluo} W. Chao, S. Luo, Z. Z. Xing and S. Zhou, arXiv:0709.1069
[hep-ph].

\bibitem{A4} E. Ma and G. Rajasekaran, \emph{Phys. Rev. D} {\bf 64}, 113012
(2001).

\bibitem{tribi} P. F. Harrison, D. H. Perkins and W. G. Scott, \emph{Phys.
Lett. B }{\bf 530}, 167 (2002); Z. Z. Xing, \emph{Phys. Lett. B}
{\bf 533}, 85 (2002); P. F. Harrison and W. G. Scott, \emph{Phys.
Lett. B} {\bf 535}, 163 (2002); X. G. He and A. Zee, \emph{Phys.
Lett. B} {\bf 560}, 87 (2003).

\bibitem{Grimus} W. Grimus and L. Lavoura, \emph{JHEP }{\bf 0011}, 042
(2000); Z. Z. Xing and S. Zhou, \emph{High Energy Phys. Nucl. Phys.
}{\bf 30}, 828 (2006).

\bibitem{CPV} E. Fernandez-Martinez, M. B. Gavela, J. Lopez-Pavon
and O. Yasuda, \emph{Phys. Lett. B} {\bf 649}, 427 (2007); Z. Z.
Xing, arXiv:0709.2220 [hep-ph].

\bibitem{antusch} S. Antusch, C. Biggio, E. Fernandez-Martinez, M. B.
Gavela and J. Lopez-Pavon, \emph{JHEP} {\bf 0610}, 084 (2006).

\bibitem{Biggio} A. Abada, C. Biggio, F. Bonnet, B. Gavela and T.
Hambye, arXiv:0707.4058 [hep-ph].

\bibitem{sen} See, e.g.,
T. Han and B. Zhang, \emph{Phys. Rev. Lett.} {\bf 97}, 171804
(2006); F. del Aguila, J. A. Aguilar-Saavedra and R. Pittau,
hep-ph/0703261.

\bibitem{pair} T. Han, B. Mukhopadhyaya, Z. Si and K. Wang,
\emph{Phys. Rev. D} {\bf 76}, 075013 (2007).

\bibitem{Smirnov} See, e.g., J. Kersten and A. Yu. Smirnov, \emph{Phys. Rev. D} {\bf 76},
073005 (2007).
\end{thebibliography}
\end{document}